\documentclass[API,prl,preprint,superscriptaddress]{revtex4}
\usepackage{graphicx}

\usepackage[usenames,dvipsnames]{color}
\usepackage{color}

\newcommand{\grbf}[1]{\mbox{\boldmath $ #1 $}}

\begin{document}



\title{Scattering properties of the three-dimensional topological insulator Sb$_2$Te$_3$:  
Coexistence of topologically trivial and non-trivial surface states with opposite spin-momentum helicity}

\author{P.\,Sessi} 
\email[corresponding author: ]{paolo.sessi@physik.uni-wuerzburg.de}
\affiliation{Physikalisches Institut, Experimentelle Physik II, 
Universit\"{a}t W\"{u}rzburg, Am Hubland, 97074 W\"{u}rzburg, Germany}
\author{O.\,Storz} 
\affiliation{Physikalisches Institut, Experimentelle Physik II, 
Universit\"{a}t W\"{u}rzburg, Am Hubland, 97074 W\"{u}rzburg, Germany}
\author{T.\,Bathon} 
\affiliation{Physikalisches Institut, Experimentelle Physik II, 
Universit\"{a}t W\"{u}rzburg, Am Hubland, 97074 W\"{u}rzburg, Germany}
\author{S.\,Wilfert} 
\affiliation{Physikalisches Institut, Experimentelle Physik II, 
Universit\"{a}t W\"{u}rzburg, Am Hubland, 97074 W\"{u}rzburg, Germany}
\author{K.\,A.\,Kokh}
\affiliation{V.S. Sobolev Institute of Geology and Mineralogy, Siberian Branch, 
	Russian Academy of Sciences, 630090 Novosibirsk,  Russia}
\affiliation{Novosibirsk State University, 630090 Novosibirsk,  Russia}
\affiliation{Saint-Petersburg State University, 198504 Saint-Petersburg, Russia}
\author{O.\,E.\,Tereshchenko}
\affiliation{Novosibirsk State University, 630090 Novosibirsk,  Russia}
\affiliation{Saint-Petersburg State University, 198504 Saint-Petersburg, Russia}
\affiliation{A.V. Rzanov Institute of Semiconductor Physics, Siberian Branch, 
	Russian Academy of Sciences, 630090 Novosibirsk, Russia}
\author{G.\,Bihlmayer} 
\affiliation{Peter Gr\"unberg Institut and Institute for Advanced Simulation, 
Forschungszentrum J\"ulich and JARA, 52428 J\"ulich, Germany}
\author{M.\,Bode}
\affiliation{Physikalisches Institut, Experimentelle Physik II, 
Universit\"{a}t W\"{u}rzburg, Am Hubland, 97074 W\"{u}rzburg, Germany}
\affiliation{Wilhelm Conrad R{\"o}ntgen-Center for Complex Material Systems (RCCM), 
Universit\"{a}t W\"{u}rzburg, Am Hubland, 97074 W\"{u}rzburg, Germany}

\date{\today}

\vspace{1cm}
\begin{abstract}
The binary chalcogenides Bi$_2$Te$_3$ and Bi$_2$Se$_3$ 
are the most widely studied topological insulators. 
Although the quantum anomalous Hall effect 
has recently been observed in magnetically doped Sb$_2$Te$_3$ 
this compound has been studied to a much lesser extend. 
Here, by using energy-resolved quasiparticle interference mapping, 
we investigate the scattering events of pristine Sb$_2$Te$_3$ surfaces.  
We find that, in addition to the Dirac fermions, another strongly spin-polarized surface resonance 
emerges at higher energies in the unoccupied electronic states. 
Although the two surface states are of different origin, i.e.\ topologically protected and trivial, respectively, 
both show strongly directional scattering properties and absence of backscattering. 
Comparison with {\em ab-initio} calculations demonstrates that this 
is a direct consequence of their spin-momentum--locked spin texture 
which is found to exhibit an opposite rotational sense for the trivial state and the Dirac state.
\end{abstract}


\maketitle


\section{Introduction}

The discovery of topological insulators (TIs) \cite{KWB2007,HQW2008,CAC2009} 
led to intense research efforts towards the potential utilization of these materials 
in spintronic, magneto-electric, or quantum computation devices \cite{FK2008,GF2010,PM2012,MLR2014}. 
In particular, the existence of a linearly dispersing, gapless surface state is of enormous interest, 
because---contrary to the trivial surface states usually found 
at surfaces of metals and semiconductors---it cannot be destroyed 
by the presence of defects and adsorbates as long as time-reversal symmetry 
is preserved \cite{CCA2010,ODZ2011,WXX2011,VPG2012,RSM2012,XNL2012,SRB2014}. 
The strong spin-orbit coupling inherent to these systems perpendicularly locks the spin to the momentum, 
creates a helical spin texture which forbids backscattering \cite{RSP2009,ZCC2009}, 
and results in spin currents that are intrinsically tied to charge currents \cite{KWB2007}. 

Among the three-dimensional (3D) TIs, the binary chalcogenides Bi$_2$Te$_3$\,\cite{CAC2009} and 
Bi$_2$Se$_3$ \cite{XQH2009} represent the most widely studied compounds. 
In contrast, the electronic properties of Sb$_2$Te$_3$\,\cite{JWC2012} 
which belongs to the same material class have been studied to a much lesser extent. 
This deficiency can directly be linked to the intrinsic strong $p$-doping characterizing the compound, 
which leads to a Dirac point that lies well above the Fermi level \cite{SBB2015,Seibel2015}. 
As a result relevant parts of the surface electronic band structure 
are inaccessible to angle-resolved photoemission spectroscopy (ARPES), 
a technique that played a key role towards the identification and study of topologically non-trivial states of matter.

Another technique frequently applied to TIs, which gives access to the scattering properties of surfaces 
not only for occupied states below but also for empty electronic states above the Fermi level, 
is quasiparticle interference (QPI) imaged with a scanning tunneling microscope (STM) 
\cite{RSP2009,ZCC2009,AAC2010,ODZ2011,SRB2014}. 
However, the observation of QPI signals in TIs depends on the deformation of the Dirac cone 
which is described by the so-called warping term \cite{F2009}. 
This contribution to the Hamilton operator leads to an effective nesting of equipotential surfaces 
and to the development of out-of-plane spin polarization components, 
which both result in new scattering channels and a higher QPI signal strength.  
Unfortunately, the warping term of Sb$_2$Te$_3$ is very weak 
as compared to Bi$_2$Te$_3$ \cite{ZCC2009,AAC2010,SOB2013}. 
This together with the above mentioned $p$-doping, 
which requires to work at energies far above the Fermi level 
and results in strongly reduced lifetimes and rapidly damped standing wave patterns, 
makes scattering experiments on Sb$_2$Te$_3$ highly challenging.


This deficiency is the more annoying as the quantum anomalous Hall effect 
has recently been observed in magnetically doped Sb$_2$Te$_3$ \cite{CZF2013,CZK2015}.  
Therefore, a comparison of the scattering properties with pristine Sb$_2$Te$_3$ would be highly interesting 
to identify the correlation between dissipation-less quantized transport 
and the onset of ferromagnetism in a topologically non-trivial material. 
Here, we close this gap by QPI of $p$-doped Sb$_2$Te$_3$ 
which contains defects that effectively scatter the surface state.  
In addition to the linearly dispersing spin-momentum--locked Dirac states, our data evidence the existence 
of another surface-related electronic feature with strongly directional scattering properties. 
Comparison with {\em ab-initio} calculations reveals that the latter originates from a trivial surface resonance 
which also possesses a helical spin texture suppressing backscattering. 
Interestingly, the helical spin texture of the Dirac state and the trivial surface resonance are opposite. 
This unique feature may allow for the independent tuning of spin and charge currents, 
for example by tuning the potentials by means of gating. 

\section{Experimental procedures}

The Sb$_2$Te$_3$ single crystal was grown 
by the modified vertical Bridgman method with rotating heat field \cite{KPK2007}.  
Stoichiometric amounts of Sb and Te were loaded to a carbon-coated quartz ampoule. 
After evacuation to $10^{-4}$ torr the ampoule translation rate and axial temperature gradient 
were set to 10\,mm/day and $\approx 15^{\circ}$/cm, respectively.  
Single crystals of about 10\,mm in diameter 
and 60\,mm in length were obtained as shown in Fig.\,\ref{fig:STS}(a). 

After growth, crystals have been cut in sizes suitable for STM experiments, 
cleaved at room temperature in ultra-high vacuum, and immediately inserted into a cryogenic STM. 
Because of the Sb$_2$Te$_3$ structure depicted in Fig.\,\ref{fig:STS}(b), 
which consist of quintuple layers that are weakly bound 
by van der Waals forces, the surface is always Te-terminated.  
Spectroscopic data have been obtained by low-temperature scanning tunneling microscopy 
at $T = 1.5 ... 4.8$\,K under ultrahigh vacuum conditions ($p \le 10^{-10}$\,mbar) 
using the lock-in technique ($f = 793$\,Hz) with a modulation voltage $U_{\rm mod} = 10 ... 20$\,meV. 
Unless otherwise stated experiments were performed at zero magnetic field.  
While maps of the differential conductance $\mathrm{d}I/\mathrm{d}U$ 
have been acquired simultaneously with topographic images in the constant-current mode,
scanning tunneling spectroscopy (STS) curves are measured by ramping the bias voltage 
after deactivation of the feedback-loop, i.e.\ at constant tip--sample separation.  

\section{Results and discussion}

\begin{figure}[t]   
\begin{minipage}[t]{0.55\textwidth}
	\includegraphics[width=0.9\columnwidth]{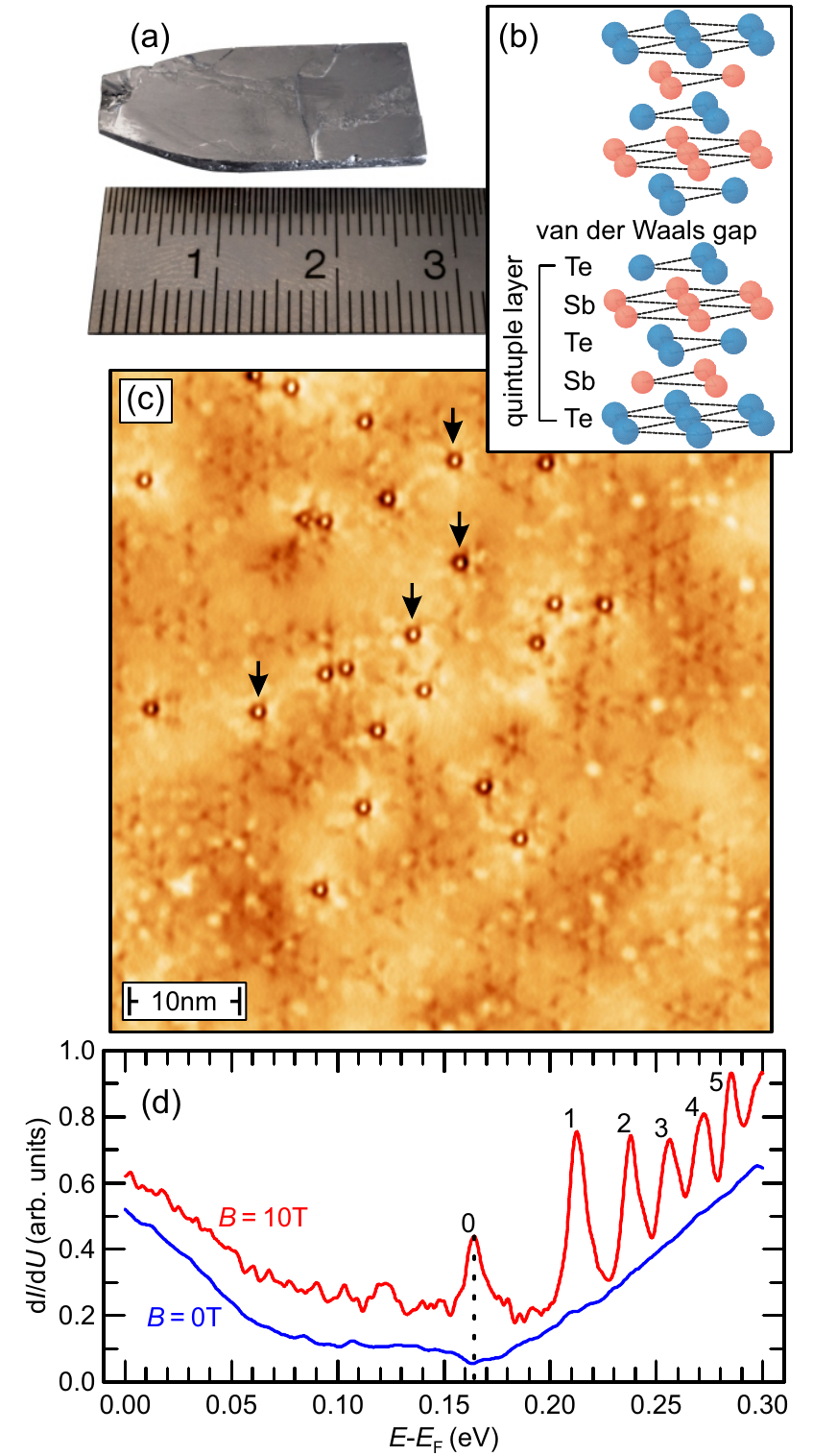}%
	\end{minipage}
	\hfill
	\begin{minipage}[b]{0.37\textwidth}
	\caption{(a) Photographic image 
	of an as-grown Sb$_2$Te$_3$ single crystal after longitudinal cleavage. 
	(b)~Sb$_2$Te$_3$ crystal structure, consisting of alternating layers of Sb and Te 
	up to the formation of a quintuple layer 
	which is weakly bound to the next one by van der Waals forces. 
	(c)~Topographic image of the Sb$_2$Te$_3$ surface (scan parameters: $I = 50$\,pA, $U = 0.5$\,V).  
	Arrows indicate the type of defects responsible for the formation of QPI patterns.
	(d)~STS data taken on Sb$_2$Te$_3$ with (red line) and without magnetic field (blue)
	[setpoint parameters: $I = 300$\,pA, $U = 0.35$\,V ($B = 10$\,T); $I = 100$\,pA, $U = 0.3$\,V ($B = 0$\,T)]. 
	In high magnetic fields we observe peaks representing Landau levels 
	whose indices are indicated by numbers.
	\label{fig:STS}}
	\end{minipage}   
\end{figure}   
Figure~\ref{fig:STS}(c) shows a topographic image of the Sb$_2$Te$_3$ surface
with defects typical for binary chalcogenides \cite{JSC2012}. 
To our experience the existence of the defect type indicated by an arrow, 
which probably corresponds to a Te surface vacancy, is crucial for the formation of a QPI pattern.
We speculate that this may be related to the fact that only this defect exhibits a potential 
that sufficiently overlaps with the wave function of the topological surface state. 
The electronic structure is investigated by analyzing the local density of states as probed by STS. 
The bottom curve of Fig.\,\ref{fig:STS}(d) shows a typical STS spectrum 
obtained on pristine Sb$_2$Te$_3$ by positioning the tip away from defects (blue line). 
The minimum corresponds to the position of the Dirac point \cite{JWC2012,JSC2012} 
which is found at about 170\,meV above the Fermi level, i.e.\ in the empty electronic states, 
confirming the intrinsic $p$-doping characterizing the material \cite{SBB2015}. 
This assignment is also corroborated by the position of the zeroth-order Landau level 
measured at a magnetic field $B = 10$\,T (red line). 

\begin{figure*}[t]   
\includegraphics[width=.99\textwidth]{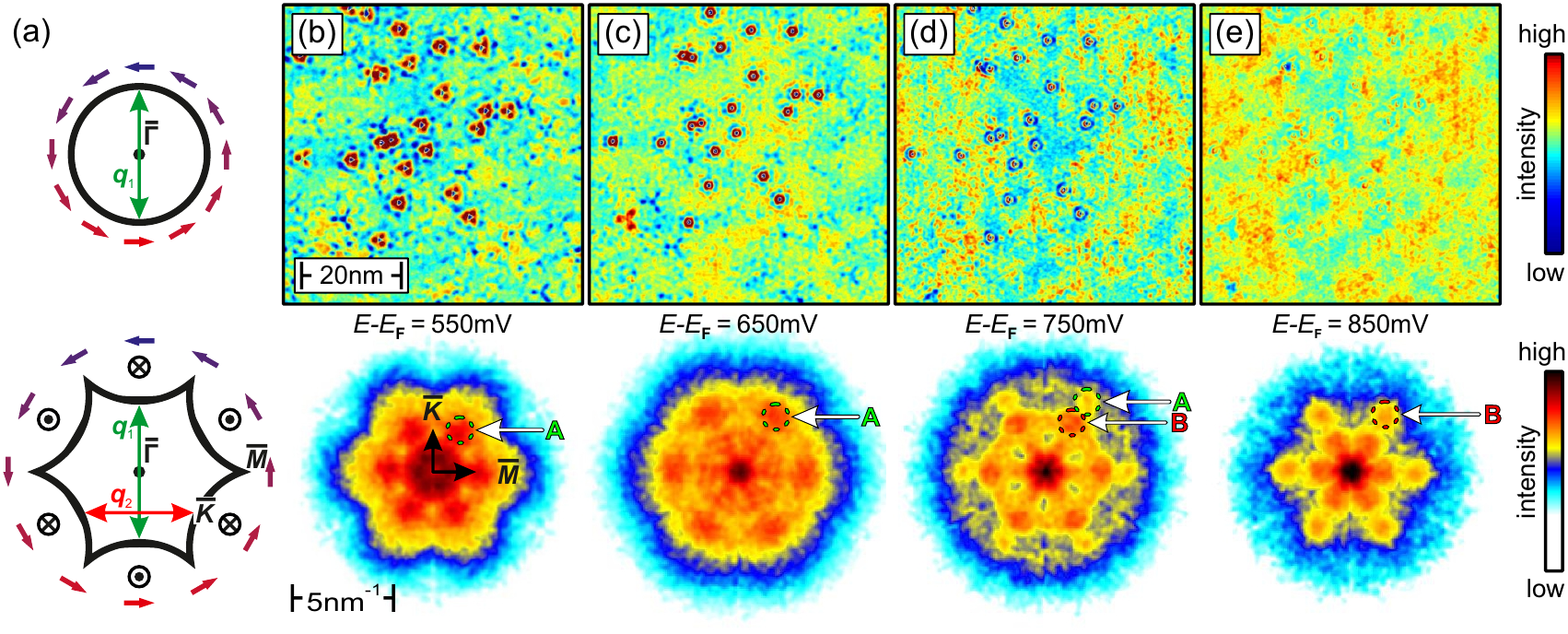}%
\caption{(a) Schematic presentation of the equipotential surface (black), the in-plane spin polarization (blue/red arrows), 
	and potential scattering vectors of a TI close to the Dirac point (upper panel) 
	and at an energy closer to the bulk conduction band (bottom).  
	Scattering vectors are indicated by ${\mathbf q_1}$ and ${\mathbf q_2}$. 
	The former is forbidden by time reversal symmetry. 
	(b)-(e) $\mathrm{d}I/\mathrm{d}U$ maps (upper panel) with corresponding FT (lower) 
	at selected representative energies obtained over the area displayed in Fig.~\protect\ref{fig:STS}(c) 
	(tunneling current $I = 50$\,pA).
	Two features labeled A and B, both located in $\overline{\Gamma M}$ directions, are visible. }	 
\label{fig:QPI}
\end{figure*}   
To investigate the scattering properties of Sb$_2$Te$_3$, we performed energy-dependent QPI experiments. 
QPI mapping makes use of the standing-wave pattern generated by elastic scattering of electronic states 
at surface defects and is known as a powerful method to study scattering mechanisms. 
While originally applied to noble metal surfaces \cite{CLE1993}, its use has been recently extended 
to investigate non-degenerate spin-polarized bands, e.g.\ on surfaces 
with strong contributions from spin-orbit coupling \cite{PhysRevLett.93.196802,KSB2013} 
and topological insulators \cite{RSP2009,ZCC2009,AAC2010,ODZ2011,SRB2014}.  
Fourier transformation (FT) translates real-space information into reciprocal space, 
thereby providing a convenient way to visualize scattering vectors, 
which correspond to points of a constant energy cut (CEC) 
connected by nesting vectors \cite{LQZ2012}. 

The shape of equipotential surfaces of TIs and the spin texture are both strongly energy-dependent. 
In Sb$_2$Te$_3$, similar to other TIs hosting a single Dirac cone 
centered around the $\overline{\Gamma}$-point of the surface Brillouin zone, 
one finds circular CECs at energies close to the Dirac point 
supporting only one nesting vector, i.e.\ backscattering (${\mathbf q_1}$). 
As schematically represented by red/blue arrows 
in the upper panel of Fig.~\ref{fig:QPI}(a), however,  
the spin polarization in this energy range is perpendicularly locked 
to the crystal momentum $k$ and completely in-plane. 
In this case backscattering is forbidden by time-reversal symmetry 
which explains the absence of any scattering channel close to the Dirac point. 

In contrast, at higher energies the introduction of the warping term 
progressively deforms the circular CEC, eventually giving rise to a snowflake-like shape, 
as shown in the lower panel of Fig.~\ref{fig:QPI}(a). 
At the same time the warping leads to an increasing out-of-plane modulation of the spin polarization.  
Although backscattering remains forbidden, it is well-known from both 
theoretical\,\cite{F2009} and experimental\,\cite{RSP2009,ZCC2009,SOB2013} investigations 
that the warping term opens new scattering channels along $\overline{\Gamma M}$ directions 
by connecting next-nearest neighbor concavely warped sides 
that are centered at the $\overline{K}$ point of the hexagram 
[schematically represented by the red arrow labeled ${\mathbf q_2}$ in Fig.~\ref{fig:QPI}(a), lower panel]. 
This scattering channel can be experimentally visualized by QPI. 
For example, Fig.~\ref{fig:QPI}(b) shows the $\mathrm{d}I/\mathrm{d}U$ map (upper panel) 
and the corresponding FT (lower panel) at an energy $E = 550$~meV above the Fermi level.  
The FT exhibits six distinct maxima that represent ${\mathbf q_2}$ (one of which is labeled A). 
With increasing energy it progressively moves away from the center of the FT,   
thereby reflecting the band dispersion relation of the Dirac cone [Fig.~\ref{fig:QPI}(c)]. 

Careful inspection of the FT of $\mathrm{d}I/\mathrm{d}U$ maps taken at higher energy 
reveals the emergence of another, unexpected scattering channel not previously observed on TIs.
This scattering channel, which is labeled B in Fig.~\ref{fig:QPI}(d), 
becomes detectable at about 750\,meV above the Fermi level. 
It also shows a well-defined directionality 
with maxima pointing along the $\overline{\Gamma M}$ direction. 
Since its length is shorter than ${\mathbf q_2}$ we can safely exclude 
an additional scattering channel involving topological states.
While it initially coexists with ${\mathbf q_2}$, the new scattering channel completely dominates 
the FT-QPI signal at higher energies, as shown in Fig.~\ref{fig:QPI}(e) at $E = 850$\,meV.  

\begin{figure}[t]   
\begin{minipage}[b]{0.51\textwidth}
	\vspace{-1cm} \includegraphics[width=.99\columnwidth]{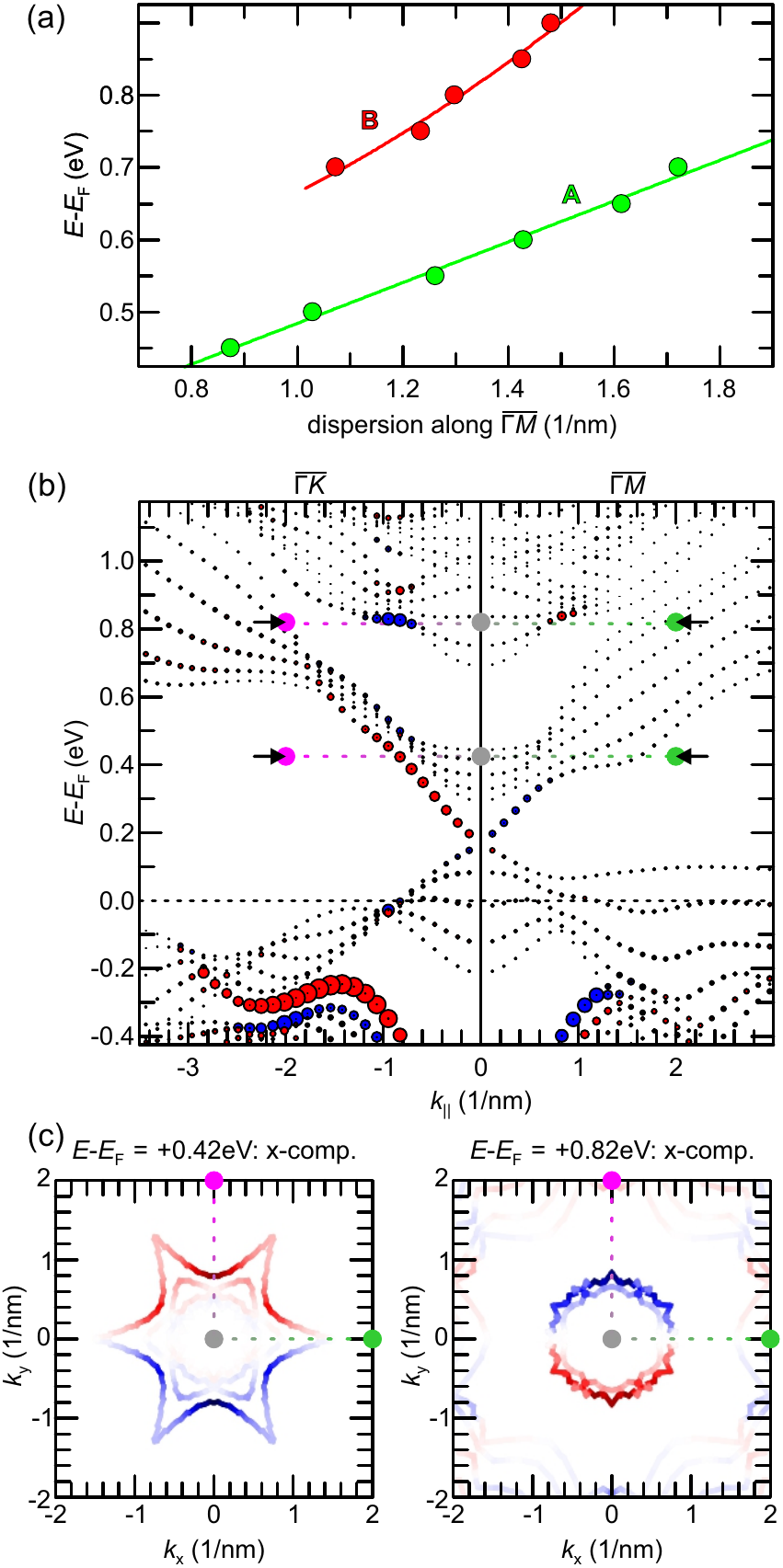}%
	\end{minipage}
	\hfill
	\begin{minipage}[b]{0.46\textwidth}
	\caption{(a) Experimentally determined energy dispersion relation 
	of features A and B [cf.\ Fig.\,\protect\ref{fig:QPI}(b)-(e)] along the $\overline{\Gamma M}$ direction. 
	Data are obtained by a quantitative analysis of FT-QPI patterns measured at different energies. 
	(b) Theoretical band structure of the Sb$_2$Te$_3$ surface. 
	Note that the Dirac point has been shifted to the experimentally determined value 
	to account for the doping-induced modification of the chemical potential (see main text for details).
	In addition to the Dirac cone another fea\-ture with an electron-like dispersion 
	and an onset energy of about 0.7\,eV above the Fermi level can be recognized. 
	(c) Constant-energy cuts taken at the energies indicated by black arrows in (b), 
	i.e.\ 0.42\,eV and 0.82\,eV above $E_{\rm F}$. 
	To better relate the dispersion relation in (b) to the constant-energy cuts of (c), 
	pink, grey, and green points are used to mark corresponding positions in the Brillouin zone 
	along the $\overline{\Gamma K}$ direction, at the $\overline{\Gamma}$ point, 
	and along the $\overline{\Gamma M}$ direction, respectively.   
	Note that the two states exhibit an opposite spin polarization.
	\label{fig:dispersion}}	 
	\end{minipage}  
\end{figure}   %
To identify the electronic states that lead to the appearance of this additional scattering channel 
on pristine Sb$_2$Te$_3$ we have quantitatively analyzed a larger series of FT-QPI maps 
that is only partially presented in Fig.~\ref{fig:QPI}(b)-(e). 
The result is shown in Fig.~\ref{fig:dispersion}(a). 
The linear dispersion relation of signal A (green dots) that represents the topological state is clearly visible. 
A fit to the data provides a Dirac point at $E_{\rm D} = (202 \pm 17)$\,meV above the Fermi level. 
This value is slightly higher than the one obtained from the STS data presented in Fig.\,\ref{fig:STS}(d),
a deviation we attribute to the warping term, which introduces a subtle deviation 
from a linear dispersion relation also visible in the calculations. 
The slope of the linear dispersion gives a Fermi velocity 
$v_{\rm F}  = (4.3 \pm 0.2) \times 10^5$\,ms$^{-1}$, 
in reasonable agreement with previous reports on binary chalcogenides \cite{JWC2012,Pauly:12.1,SRB2014}.
In contrast, scattering channel B (red dots) appears in a narrow energy and crystal momentum range 
around $E - E_{\rm F} = 800$\,meV and $k_{\rm exp} = 1.3$\,nm$^{-1}$ only and cannot be reasonably fit.  
Correspondingly, the red line serves as a guide to the eye only.  

These experimental data have been compared with {\em ab-initio} DFT calculations
(obtained as described in Ref.~[\onlinecite{Pauly:12.1}]).  
We would like to emphasize, that DFT calculations are always performed for ideal crystals and, 
therefore, always result in a Fermi level positioned at or near the valence band maximum (VBM) 
which is close to the Dirac point (DP) here.
For this reason the experimentally observed doping-induced shift 
of the chemical potential cannot be simulated in DFT.  
Any comparison between the experiment and theory must be performed on a relative level, 
i.e.\ by comparing the position of electronic bands with respect to certain reference points.  
In our experiments the Fermi level is unambiguously defined by the zero bias condition in STS ($U = 0$\,V).  
Furthermore, the energetic position of the Dirac point relative to the Fermi level 
can be determined with an accuracy we estimate to about $\pm 10$\,meV.  
Therefore, we have chosen to shift the theoretical Dirac point energy 
to the experimental value, i.e. $+170$\,meV [cf.\ Fig.\,\ref{fig:STS}(d)].   

The band structure obtained from DFT is reported in Fig.~\ref{fig:dispersion}(b). 
In good agreement with the experimental data presented in Fig.~\ref{fig:dispersion}(a),
DFT evidences the existence of an additional surface resonance
with a charge density centered around $E - E_{\rm F} = 800$\,meV, 
but at a slightly lower value of the crystal momentum, $k_{\rm theo} \approx 1.0$\,nm$^{-1}$. 
To shed light on the strongly directional character of the scattering events involving from this band, 
its spin texture has been calculated and compared with the topological state. 
Results for the spin polarization component along the $x$-direction is reported in Fig.~\ref{fig:dispersion}(c). 
Interestingly, the two electronic states under discussion here, which are energetically located 
at around 0.42~eV and 0.82~eV above the Fermi level, exhibit opposite spin helicities.
Furthermore, a detailed analysis of the other directional components of the spin polarization (not shown here) 
reveals a helical texture with significant out-of-plane components. 
This is expected for the warped Dirac-cone, but it is also observed for the higher lying state. 

\begin{figure*}[t]   
\includegraphics[width=.7\textwidth]{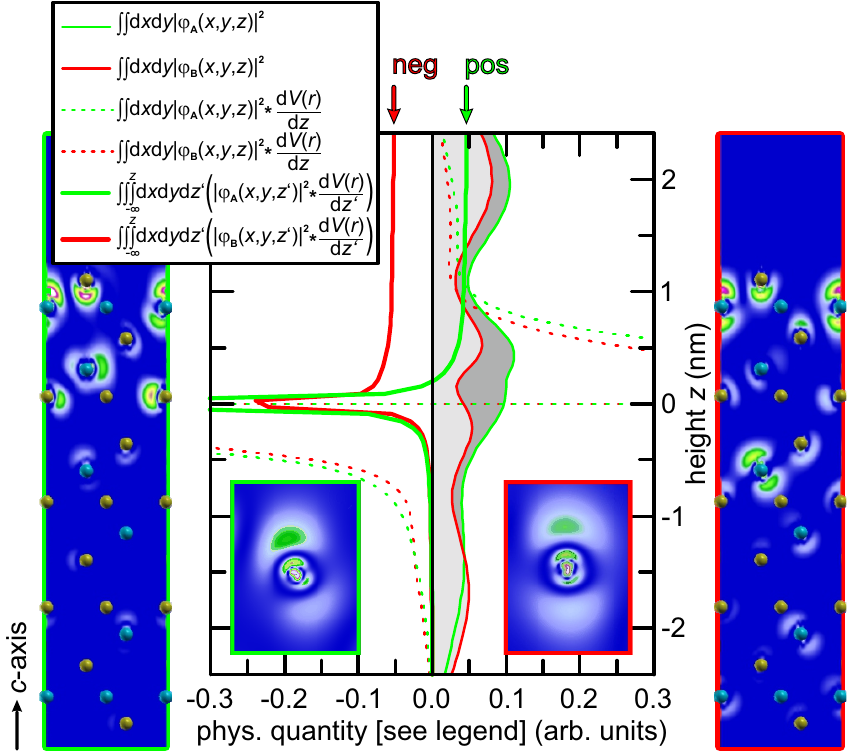}%
\caption{Charge density plots in the $y$--$z$-plane of the surface states 
	located at a $k$-point at 1/nm in $\overline{\Gamma K}$ direction 
	[$E - E_{\rm F} = 0.42$~eV (left panel) and 0.82~eV (right panel)]. 
	Close-up views of the states are shown in the central panel as insets 
	(framed green and red for the lower and higher lying electronic state, respectively). 
	In-plane--averaged charge densities 
	($\int \!\!\! \int \!\! {\rm d}x {\rm d}y | \phi(x,y,z) |^2$ around the subsurface Sb atom at $z=0$) 
	are plotted as green and red lines in the middle panel. 
	Moreover, the product of charge density with the derivative of a $1/|r|$ potential (dotted lines)
	and the integral of this product in $z$-direction as function of the integration region $[-\infty, z]$ are shown. 
	Note that sign of the latter quantity converges for large $z$ to a positive value
	for the lower lying topological surface state but to a negative value for the surface resonance, 
	giving rise to the opposite rotational sense of their spin textures.  
	 \label{fig:states} }	 
\label{Figure4}
\end{figure*}   
To understand this surprising result we analyze the charge density in more detail, 
following the model of Nagano {\em et al.}~\cite{Nagano:09.1}.
This model unfolded that the major effect of spin-orbit coupling, 
that determines the spin orientation, originates from a region close to the nucleus 
where the potential gradient $V(r)$ is strongest~\cite{Bihlmayer:06.1}. 
The charge density plots shown in Fig.~\ref{fig:states} reveal that both states 
are predominantly localized at subsurface Sb atoms and exhibit $p_z$ character. 
In spite of their similar shape, the strength of the spin-orbit coupling and the resulting spin polarization 
of both electronic features is found to sensitively depend 
on subtle variations of the exact spatial distribution of the charge density.  
If the state is located on average a bit above the position of the heavy Sb nucleus 
(i.e.\ shifted to positive $z$-direction) it will be influenced by regions where $\partial_z V$ is positive,
giving rise to an effective electric field acting on the state. 
Due to the $\grbf{\sigma} \cdot (\mathbf{k} \times \nabla V)$ term of the spin-orbit coupling, 
this favors one particular spin direction for a given $\mathbf{k}$ vector 
leading to a certain helicity of the spin texture. 
If the state is shifted a bit in the negative $z$-direction, 
however, the other spin direction (and helicity) will be favored. 
In the middle panel of Fig.~\ref{fig:states}, the charge densities of the states at 0.42~eV and 0.82~eV, 
both in-plane--averaged around the subsurface Bi atoms, 
are shown as thin black and red lines, respectively.
It appears that the state at 0.42~eV is a bit more displaced towards the vacuum. 
To quantify the effect, we multiply the charge density 
with the $z$-derivative of an atomic ($1/|r|$) potential 
and integrate the product (dotted line in the middle panel of Fig.~\ref{fig:states}) in the $z$-direction. 
Indeed, the integrals (thick full lines) converge to values of opposite sign for large $z$, 
thereby confirming our hypothesis that spin-orbit coupling arising from this Sb atom 
induces opposite spin orientations of the two states.

\section{Summary}
In conclusion, we have investigated the scattering properties 
of the three-dimensional topological insulator Sb$_2$Te$_3$. 
While no scattering events occur close to the Dirac point, 
scattering vectors connecting next-nearest neighbor concavely warped sides 
of the hexagram-shaped constant energy cuts appear as soon as the warping term 
introduces a substantial out-of-plane spin polarization. 
A new surface-related electronic band with strongly directional scattering channels 
has been revealed at higher energy. 
Although its nature is not topological, it shows a well-defined spin texture, 
with a spin-momentum locking opposite to the Dirac fermion. 

\section{Acknowledgements}
This work was supported by the Deutsche Forschungsgemeinschaft 
within SPP 1666 (Grant Nos. BO1468/21-1 and BI). 
K.A.K.\ and O.E.T.\ acknowledge the financial support 
by the RFBR (Grant nos.\ 14-08-31110 and 15-02-01797 )
and Petersburg State University (project no.\ 11.50.202.2015).


\end{document}